\documentclass[twocolumn]{aastex62}
\submitjournal{ApJ}

\usepackage[utf8]{inputenc}
\usepackage{amsmath}
\usepackage{newtxtext,newtxmath}
\usepackage{bm}



\shorttitle{Resonant relaxation in globular clusters}
\shortauthors{Meiron \& Kocsis}

\begin{document}

\title{RESONANT RELAXATION IN GLOBULAR CLUSTERS}

\author[0000-0003-3518-5183]{Yohai Meiron}
\affil{Institute of Physics, E\"otv\"os University, P\'azm\'any P. s. 1/A, Budapest, 1117, Hungary}

\author[0000-0002-4865-7517]{Bence Kocsis}
\affil{Institute of Physics, E\"otv\"os University, P\'azm\'any P. s. 1/A, Budapest, 1117, Hungary}

\begin{abstract}
Resonant relaxation has been discussed as an efficient process that changes the angular momenta of stars orbiting around a central supermassive black hole due to the fluctuating gravitational field of the stellar cluster. Other spherical stellar systems, such as globular clusters, exhibit a restricted form of this effect where enhanced relaxation rate only occurs in the directions of the angular momentum vectors, but not in their magnitudes; this is called vector resonant relaxation (VRR). To explore this effect, we performed a large set of direct N-body simulations, with up to 512k particles and \textasciitilde{}500 dynamical times. Contrasting our simulations with Spitzer-style Monte Carlo simulations, that by design only exhibit 2-body relaxation, we show that the temporal behavior of the angular momentum vectors in $N$-body simulations cannot be explained by 2-body relaxation alone. VRR operates efficiently in globular clusters with $N>10^4$. The fact that VRR operates in globular clusters may open way to use powerful tools in statistical physics for their description. In particular, since the distribution of orbital planes relaxes much more rapidly than the distribution of the magnitude of angular momentum and the radial action, the relaxation process reaches an internal statistical equilibrium in the corresponding part of phase space while the whole cluster is generally out of equilibrium, in a state of quenched disorder. We point out the need to include effects of VRR in Monte Carlo simulations of globular clusters.
\end{abstract}

\keywords{globular clusters --  stars: kinematics and dynamics}

\section{Introduction\label{sec:introduction}}

In stellar systems such as globular clusters and nuclear stellar clusters, gravitational encounters provide a mechanism for the exchange of orbital energies and angular momenta. This process is called \emph{2-body relaxation} and is well described by Chandrasekhar's scattering theory (\citealt{Chandrasekhar42} and references thereafter) as a diffusion process. The main assumptions made in this theory are that the velocities exhibit Brownian motion independently due to the superposition of a large number of independent instantaneous 2-body encounters within a uniform medium. This process causes the energy distribution to diffuse. Numerical studies have shown that the energy diffusion rate is well described by this theory even in realistic self-gravitating stellar systems which are inhomogeneous and where the interactions are non-local and non-instantaneous, once the free parameter of the model, the Coulomb logarithm, is fitted \citep[e.g.,][]{Giersz94}. Applying this analytical theory to globular clusters can match the simulated diffusion rate of energy and angular momentum magnitude, the so-called relaxation timescale \citep[e.g.,][]{Aarseth75}.

\citet[henceforth RT96]{Rauch+96} demonstrated that this kind of gravitational encounters does not fully describe relaxation in stellar systems. Focusing on spherical systems dominated by a central mass, such as galactic nuclei, where the stars' orbits are nearly Keplerian, they showed that there is an enhanced rate of angular momentum relaxation compared to what is predicted from Chandrasekhar's scattering theory. This enhanced relaxation rate is caused by persistent torques among the Keplerian orbits, which act coherently until the orbits are sufficiently perturbed or until they precess significantly. In this case, the Keplerian orbits can be thought of as ellipse-shaped wires; each pair of such wires exerts mutual gravitational torques, thereby exchanging angular momenta. This change is coherent over timescales shorter than the precession time in the stellar cluster (either the mass precession timescale or general relativity precession; RT96). On longer timescales a random walk-like behavior takes place where the effective step size is set by the coherence time, which is much longer than the orbital time.

This process, called \emph{resonant relaxation}\footnote{\citet[section 5]{Henon59} already speculated about resonant relaxation, but his idea was very different from the modern concept and had to do with enhanced \emph{energy} exchange among stars with common orbital period. Hénon hypothesized that this kind of process will lead the mass distribution of any globular cluster to evolve toward the isochrone model. However it is now known not to be the case.}, is not restricted to Keplerian orbits. As also noted by RT96, spherical systems not dominated by a central mass (such as globular clusters) also exhibit a more restricted form of the phenomenon. In a general spherical potential, orbits are still restricted to motion in a plane, but the radial and azimuthal orbital frequencies are no longer equal, and orbits precess generally on timescales comparable to the orbital time. These orbits look like ``rosettes'' that generally do not close and thus eventually sample all points between the pericenter and apocenter, and can be thought of as annuli. A pair of such annuli exerts mutual torques. In this case, the component of the torque vector parallel to the angular momentum is zero, meaning that the angular momentum magnitudes are left unchanged. Thus, the angular momentum magnitudes do not relax. The perpendicular component is nonzero, and therefore the orientations of the angular momentum vectors do relax. This process is called \emph{vector resonant relaxation}
(VRR) to distinguish it from relaxation of the angular momentum magnitudes in a Keplerian potential or \emph{scalar resonant relaxation} (SRR). Although similar torques drive both SRR and VRR, the latter is much more efficient because the coherence time of the former is much shorter due to precession \citep{Hopman+06,BarOr16,Alexander17}.

More generally, resonant relaxation arises if the smooth component of the gravitational potential admits action-angle variables, where its fundamental frequencies satisfy a relation $n_1\Omega_1 + n_2 \Omega_2 + n_3 \Omega_3 = 0$ with $n_{1}$, $n_{2}$, $n_{3}$ not all zero, integer coefficients (this is called the resonance criterion, see also \citealt{Merritt13}). During resonant relaxation, the energies of orbits in the smooth potential are approximately conserved, but the mutual interactions drive the actions, that correspond to the resonance criterion, to change rapidly. In the case of stars moving in a Keplerian potential, the resonance condition may be satisfied by any $n_{2,3}$ integer as $\Omega_2=\Omega_3=0$ for the argument of pericenter and ascending node, which leads to the rapid change of the angular momentum (RT96).

In both the Keplerian and general spherical potentials, this process does not cause energy change because the orbitally-averaged structures are symmetric. Thus, the rate of energy change is dominated only by gravitational encounters or 2-body relaxation.  Since the relaxation of energy is nonresonant (i.e. occurs due to 2-body encounters), the energy of any particular star performs a random walk, which means that on average the square change of energy grows linearly with time:
\begin{equation}
\mathrm{rms}\,(\Delta E)\propto\sqrt{t}.
\end{equation}
This occurs on timescales longer than the 2-body coherence time but short enough that the star is not transported too far in energy space. The 2-body coherence time is a fraction of the dynamical time of the system and generally shorter than the timescales of interest. The angular momentum magnitude is similarly influenced by 2-body encounters, but SRR (in the case of nearly Keplerian orbits) imposes additional stochasticity with coherence time that equals the mass precession time, so on timescales longer than the dynamical time and shorter than the precession time,
\begin{equation}
\mathrm{rms}\,(\Delta L)\propto\eta_{\mathrm{s}}\sqrt{t}+\beta_{\mathrm{s}}t,
\end{equation}
where $\eta_{\mathrm{s}}$ and $\beta_{\mathrm{s}}$ are parameters introduced by RT96 to quantify the relative strength of the two processes. Finally, the angular momentum direction will change in a coherent way as long as there exist a component of the torques that is temporally correlated. Indeed, in a spherical potential orbits are planar, implying that the angular momentum vector directions execute a random walk on the sphere on timescales longer than the typical reorientation timescale. The coherent evolution of the angular momentum directions (i.e. the orbital planes) is self-quenching because as the orbital planes pivot, so do the torques they generate. Thus, the VRR timescale is itself the step size for the random walk of the angular momentum direction.\footnote{This is under the assumption that VRR is more efficient in randomizing the angular momentum directions than 2-body relaxation; this is not trivially the case but see Section \ref{sec:discussion}. (if 2-body time > VRR time then we might not see correlated behavior at all, this is expressed by $\eta$ and $\beta$).} So on timescales shorter than the coherence time of VRR, this process can be similarly parametrized\footnote{\label{ft:sphere}This parametrization breaks down due to the fact that the random walk is on a sphere, once the angular displacement of $\Delta\bm{L}$ is non-negligible \citep[see][for a better suited description]{Kocsis+15}.} as
\begin{equation}
\mathrm{rms}\,(|\Delta\bm{L}|)\propto\eta_{\mathrm{v}}\sqrt{t}+\beta_{\mathrm{v}}t.
\end{equation}

Numerical studies of resonant relaxation extend back to RT96 who utilized two approaches to investigate the problem. They performed $N$-body simulations under some restricted conditions and also $N$-wire simulations. In the former method, each star is represented by a particle, and its equations of motion are integrated such that the six phase-space coordinates are known at each time. In the latter method orbital averaging is performed such that stars are represented by ellipse-shaped wires, with the mass distributed on the wire in proportion to the time spent there by the star during its orbit. Their $N$-body simulations had a smooth background potential, in which ``background'' particles orbited, providing the torque to additional ``test'' particles for which the energies and angular momenta were followed to measure the relaxation effects. From the simulation results, they were able to show the expected correlated behavior of the angular momentum and that resonant relaxation is effective in changing the stars' angular momenta (magnitude and direction) when the background potential is Keplerian, and the angular momentum directions when the background is the isochrone potential. They also estimated the coefficients $\eta_{\mathrm{s,v}}$ and $\beta_{\mathrm{s,v}}$ for these two cases. This was improved by \citet{Rauch+98}, who performed a similarly restricted $N$-body simulation to explore the enhancement of tidal disruption rate due to SRR.

The $N$-wire method (also referred to as Gauss's method, see \citealt{Touma+09}) is an approximate method that isolates the effects of resonant relaxation. While energies cannot exchange between the wires, the angular momentum evolution can potentially be followed for much longer times. In the case of RT96, because of close encounters between the wires dominating the computation time, the $N$-wire simulations were actually significantly slower than the $N$-body simulations. They do report however that the results from the two methods agree qualitatively.

RT96 ignored the dependence on the orbital elements. \citet{Hopman+06}
used theoretical arguments for the dependence of the resonant relaxation timescales on energy. In order to obtain the steady state distribution of stars (in a single-mass population) around a supermassive black hole, they constructed a Fokker--Planck model (in energy) accounting for the sink (i.e. tidal disruption or accretion of stars onto the black hole) by SRR, and calibrated it using the coefficients measured by RT96. \citet{Gurkan+07}, also explored the dependence of the effect on eccentricity by calculating the global torque exerted on a test star (i.e. test orbit) by a large number of static elliptical wires representing the nuclear stellar cluster.

\citet{Eilon+09} performed a larger and more self-consistent set of $N$-body simulations. Although using at most only 200 particles, all mutual interactions were accounted for (rather than using non-interacting background particles). Additionally, they used an unsoftened gravitational potential with close encounters regularized using the KS method \citep{Kustaanheimo65}. They too measured the $\eta_{\mathrm{s,v}}$ and $\beta_{\mathrm{s,v}}$ coefficients from the simulation results. They found that $\eta_{\mathrm{s,v}}$ were a factor of $\sim3$ bigger than obtained by RT96 which was attributed to the large softening used by the latter; they also found $\beta_{\mathrm{s}}$ to be a factor of $\sim2$ bigger, but the source of this discrepancy was not identified. However, their results were consistent with the static wire experiments performed by \citet{Gurkan+07}.

\citet{Kocsis+15} presented a new method to specifically explore VRR. Their method is based on the assumption that the rate of precession is much faster than the rate at which the orbital planes change their orientation. Orbital averaging is performed over the entire precession cycle such that stars are represented by annuli, with the mass distributed on the surface in proportion to the time spent there by the star during its orbit, thus, it was called $N$-ring. With this method they measured VRR not only in the temporally coherent regime as previous works but also in the random-walk regime. For the coherent regime, they found a factor $\sim 3$ slower relaxation than \citet{Eilon+09} or the results of RT96 for a Keplerian background potential. The same factor $\sim 3$ slower relaxation rate was also found earlier by RT96 for the isochrone background potential which drives rapid apsidal precession.

These previous authors aimed most of their attention at the galactic nucleus or Kepler potential problem. Orbits in globular clusters precess quickly compared to their orbital periods, so SRR is not expected to occur. VRR, on the other hand, is expected to operate in globular clusters if one observes the scalings presented in RT96, and crudely replaces the central point mass $M$ by the mass of the cluster $mN$ (where $N$ is the number of stars, $m$ is their mean mass); this gives a 2-body relaxation time $\sim N P$ and VRR time $\sim \sqrt{N} P$, where $P$ is the orbital period. Therefore, the latter should be $\sqrt{N}$ shorter than the former, indicating that for $N\sim 10^4$--$10^6$, VRR is expected to operate at a rate 100--1000 faster than two-body relaxation time and be the primary cause for orbital plane reorientation in these systems.

In this paper we consider the efficacy of VRR in globular clusters as well as similar systems, such as spherical dwarf galaxies. We perform direct $N$-body simulations of single-mass globular clusters that follow the \citet{Plummer11} mass distribution. In our largest simulation, the number of particles was $N=512\mathrm{k}$ (where $\mathrm{k=1024}$). The large number of particles in the simulations, which is realistic for many globular clusters, is needed to clearly differentiate the effects of 2-body relaxation from VRR.

The fact that VRR operates in globular clusters may open way to use powerful tools in statistical physics for their description. In particular, since the distribution of orbital planes relaxes much more rapidly than the distribution of the magnitude of angular momentum and the radial action, the relaxation process reaches an internal statistical equilibrium in the corresponding part of phase space while the whole cluster is generally out of equilibrium, in a state of quenched disorder. In this case, statistical mechanics may be utilized to understand the long-term behavior of the system \citep{Roupas+17,Takacs+18}. Curiously, \citet{Kocsis+11} as well as \citet{Kocsis+15} have shown that the Hamiltonian of VRR in various limits is reminiscent of that of various models in condensed matter physics, particularly the $N$-vector model, liquid crystals, and point vortices on the sphere; which leads to similarities in their thermodynamic behavior \citep{Roupas+17,Takacs+18}. At zero temperature, the ground state of VRR is an ordered state of aligned orbits (parallel or antiparallel) which represents a disk in which stars orbit in either sense. At a nonzero temperature, the system undergoes a first order phase transition in the canonical ensemble between the aligned ordered phase and the spherically distributed disordered phase. A gravitating system with a nonzero rotation is analogous to a liquid crystal in a nonzero external magnetic field. In both cases, the system admits a critical value of angular momentum or magnetic field, at which the phase transition becomes second order, and at higher values there is a smooth crossover. The system also admits stable statistical equilibria with negative absolute temperature, a curious phenomenon in statistical physics \citep{Braun52,2015AmJPh..83..163F,Dunkel:2013fha,PhysRevE.91.052147,2015JSMTE..12..002C,PhysRevE.93.032149}. The correspondence between these different fields of physics may possibly have far reaching interdisciplinary implications.

Another interesting implication of VRR is that in multimass anisotropic systems, heavier and lighter objects are expected to decrease and increase their inclination in VRR equilibrium, respectively. This process is similar to dynamical friction caused by 2-body relaxation, in which the velocity dispersion of heavier objects is reduced to approach energy-equipartition. Since as we show in this paper, VRR operates in globular clusters, resonant dynamical friction reduces the dispersion in orbital inclinations for heavier objects. This leads to the formation of a disk of massive stars and black holes in galactic nuclei \citep{Szolgyen+18}, which may have far reaching implications for the dynamics of globular clusters (Sz\"olgyen, Meiron, and Kocsis, in prep).

This paper is organized as follows. In Section \ref{sec:simulations} we discuss the $N$-body simulations and the method to measure the rates of 2-body relaxation and VRR from the data. In Section \ref{sec:2body} we compare our results for 2-body relaxation only to scattering theory and a Monte Carlo $N$-body code. In Section \ref{sec:discussion} we discuss the implications of the simulation results to real globular clusters and spherical dwarf galaxies.

\section{Direct simulations}\label{sec:simulations}

To examine VRR in globular clusters, we ran a series of $N$-body simulations using the phiGRAPE code \citep{Harfst+07}, a direct-summation $N$-body code that uses the Hermite integration scheme with block timesteps \citep{Makino91}. The initial conditions are a Plummer model with a mass and virial radius of unity as expressed in Hénon units.\footnote{\label{fn:units}Also known as $N$-body units, In this unit system, the mass unit is set to the total mass of the cluster $M$ and the length unit is its virial radius $R$. The time unit is $[T]=\sqrt{R^3/(GM)}$, the energy and angular momentum units are $[E]=GM/R$ and $[L]=\sqrt{GMR}$, respectively. The virial radius relates to the Plummer radius $r_0$ through the relation $R=\frac{16}{3\pi}r_0$.} We performed two simulations with $N=128\mathrm{k}$ particles differing by the random seed, one simulation with $N=256\mathrm{k}$ and one simulation with $N=512\mathrm{k}$. The gravitational interactions are softened with softening length of $3\times10^{-4}$ length units to ensure that binary systems do not form. Of the four models presented, three ran up to 2000 time units, and one (the largest simulation with $N=512\mathrm{k}$) ran up to 4000 time units. These times were chosen because they are sufficiently long to observe 2-body relaxation and VRR in the examined phase-space regions of the simulations, but not too long that cluster evolution plays a role.

\begin{table}
\begin{center}
\begin{tabular}{|c|c|c|c|c|c|c|}
\hline Region & $E$ & $L$ & $\Delta E$ & $\Delta L$ & $a$ & $e$\\
\hline \hline (I) & $-1.21$ & $0.16$ & $0.05$ & $0.05$ & $0.2$ & $0.5$\\
\hline (II) & $-0.78$ & $0.41$ & $0.05$ & $0.05$ & $0.4$ & $0.5$\\
\hline (III) & $-0.30$ & $0.30$ & $0.15$ & $0.15$ & $1.0$ & $0.9$\\
\hline (IV) & $-0.30$ & $1.20$ & $0.15$ & $0.15$ & $1.0$ & $\lesssim0.3$\\
\hline
\end{tabular}
\end{center}
\caption{Four representative initial regions of orbits in $(E,L)$-space. The corresponding semi-major axis $a$ and eccentricity $e$ for the $(E,L)$ values are given to one digit accuracy, while in the the case of region (IV) the spread of eccentricities is larger than in the other regions. $E$, $L$, and a are in Hénon units for a Plummer model with virial radius of one (see text for details).\label{tab:regions} }

\end{table}

\begin{figure}
\includegraphics[width=1\columnwidth]{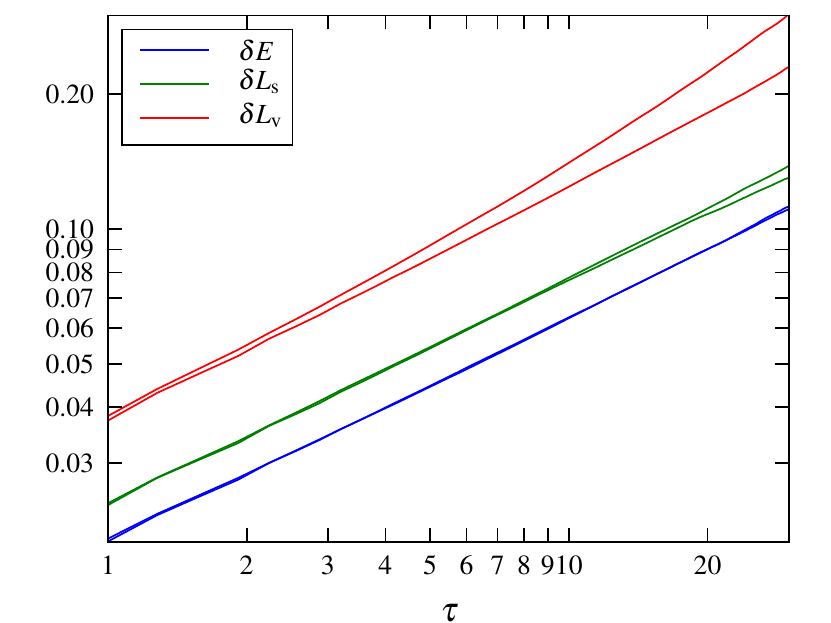}
\caption{The root-mean-square (rms) relative changes in the energy (blue) and angular momentum (magnitude in green, direction in red) as a function of time (normalized by orbital period) for two simulations of a Plummer sphere with $N=128\mathrm{k}$ particles, differing only by random seed. The lines shown are the rms of the relative changes of all stars in Region II (``most typical'' energies and angular momenta). The difference between the two red curves indicates that the coherent rate of angular momentum change is different for the two initial conditions, as expected. \label{fig:delta-example}}
\end{figure}

\begin{figure}
\includegraphics[width=1\columnwidth]{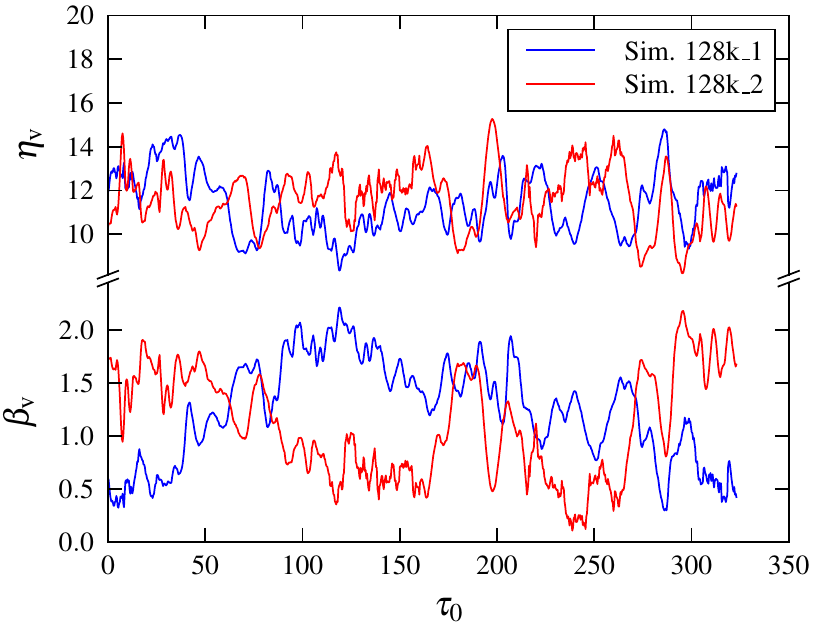}
\caption{Variation in the dimensionless fitting parameters $\eta_{\mathrm{v}}$ and $\beta_{\mathrm{v}}$ for Region II as a function of initial time $\tau_0\equiv t_0/P$ for two simulations with $N=128\mathrm{k}$. Coherent VRR is represented by $\beta_{\mathrm{v}}$ which fluctuates greatly throughout the simulation and between the two simulations, indicating that the global torques vary significantly. Note the anticorrelation between $\eta_{\mathrm{v}}$ and $\beta_{\mathrm{v}}$, the correlation coefficient between their $\tau_0$-derivatives is $\approx -0.99$. \label{fig:eta-beta}}
\end{figure}

We measured the dimensionless coefficients in a similar way to RT96 and \citet{Eilon+09}. We study the rate of change of energy and the size and direction of angular momentum in four representative regions similar to \citet[Paper I]{Meiron+18} as listed in Table \ref{tab:regions}. For each region, Table \ref{tab:regions} specifies the midpoint and width of the region in specific energy and specific angular momentum $(E, L)$. We tag all particles that are within this region initially\footnote{Because of 2-body relaxation, particles may come in and out of that region, but as long as the cluster's 6D phase space distribution function (\textsc{df}) does not evolve significantly, the number of particles there is constant in time within some statistical fluctuations. In the 512k run, each region had between 14k and 16k particles, with poissonian fluctuations.} (at $t=t_{0}$). We record each particle's $E$ and $\bm{L}$ at every subsequent step and define
\begin{align}\label{eq:delta E}
\delta E & =\frac{E-E_{0}}{E_{0}}\\\label{eq:delta L}
\delta L_{\mathrm{s}} & =\frac{|\bm{L}|-|\bm{L}_{0}|}{L_{\mathrm{c}}}\\
\delta L_{\mathrm{v}} & =\frac{|\bm{L}-\bm{L}_{0}|}{L_{\mathrm{c}}}
\end{align}
where $E_{0}$ and $\bm{L}_{0}$ correspond to $t=t_{0}$, and $L_{\mathrm{c}}$ is the circular angular momentum corresponding to $E_{0}$. For each region, the temporal behavior of these quantities is modeled as follows following RT96:
\begin{align}
\mathrm{rms}\,(\delta E) & =\alpha \frac mM\sqrt{N}\sqrt{\tau}\label{eq:rms-deltaE}\\
\mathrm{rms}\,(\delta L_{\mathrm{s}}) & =\eta_{\mathrm{s}}\frac mM \sqrt{N}\sqrt{\tau}\label{eq:rms-deltaLs}\\
\mathrm{rms}\,(\delta L_{\mathrm{v}}) & = \frac mM\sqrt{N}(\eta_{\mathrm{v}}\sqrt{\tau}+\beta_{\mathrm{v}}\tau)\label{eq:rms-deltaLv}
\end{align}
where rms denotes the root-mean-square of the selected particles (that are initially in the given region), $N$ denotes the number of particles in the full cluster, i.e., $M=Nm$, and $\tau$ is a dimensionless time
\begin{equation}
\tau \equiv \frac{t-t_{0}}{P}
\end{equation}
where $P$ is the orbital period corresponding to the center of the region in $(E,L)$-space. Here $\alpha$, $\eta_{\rm s}$, $\eta_{\rm v}$, and $\beta_{\rm v}$ are dimensionless fitting parameters which we determine for each region separately. Note that this model is only appropriate in a statistical sense, the fitting parameters may fluctuate depending on the initial condition and $t_0$, and that it is also only valid when the rms quantities are much smaller than unity. Examples of the time dependence of the rms changes in $E$, $L$, and $\bm{L}$ in two simulations are shown in Figure~\ref{fig:delta-example}, where the $t_0$ dependence of the fitting parameters is shown in Figure~\ref{fig:eta-beta}. Figure~\ref{fig:eta-beta} shows that the rate of change during the coherent phase of VRR varies significantly among the two 128k simulations. More generally, we expect $\delta L_{\rm v}$ and $\beta_{\rm v}$ to vary with $t_0$ and for different initial conditions because the rate at which orbital planes reorient is set by the instantaneous fluctuating $\sqrt{N}$ component of the gravitational field, which is mostly encoded in its quadrupolar moment, a global characteristic of the cluster. Therefore the torques experienced by the members of the cluster, which set $\delta L_{\rm v}$ and $\beta_{\rm v}$, are expected to be correlated \citep{Kocsis+11}.

The fitting parameters depend on the assumed beginning and end times of the time series, $\tau_1 \leq \tau \leq \tau_2$, which must be specified carefully. Random walk behavior for 2-body relaxation, leading to the $\sqrt{\tau}$ terms in Equations~\eqref{eq:rms-deltaE}--\eqref{eq:rms-deltaLv}, is valid only beyond its correlation time which is roughly the orbital period, implying that $\tau_1\gg 1$ (RT96). Furthermore, $\tau_2$ must be sufficiently small that the selected stars have not moved significantly in $(E,L)$-space, and experience a relaxation rate characteristic for their region (see Paper I). Finally, the adopted parametrization of VRR is only valid for small angular shifts of $\bm{L}$ (see footnote \ref{ft:sphere}), which means that $\delta L_{\rm v}\ll 1$ must be satisfied for all stars in the sample, which also sets an upper bound on $\tau_2$. For the fitting of both $\delta L_{\mathrm{s}}$ and $\delta E$ we chose $\tau_{1}=1$ and $\tau_{2}$ was set such that the value of $\mathrm{rms}[\delta E(\tau)]$ or $\mathrm{rms}[\delta L_{\mathrm{s}}(\tau)]$, whichever was fitted, was $\leq0.20$. For the fitting of ${\rm rms}[\delta L_{\mathrm{v}}(\tau)]$, we chose $\tau_{1}=5$ after observing that the fit values for $\beta_{\mathrm{v}}$ do not change significantly when $\tau_{1}$ is changed in the range between 1 and 20. The choice of $\tau_{2}$ was set such that $\mathrm{rms}[\delta L_{\mathrm{v}}(\tau)] \leq 0.25$ in the fit interval. Figure \ref{fig:eta-beta}
shows the values of $\eta_{\mathrm{v}}$ and $\beta_{\mathrm{v}}$ in Region II as functions of $t_{0}$ for the two simulations with $N=128\mathrm{k}$.

Table \ref{tab:results-N} shows the average values of the fitting parameters for all simulations. In both simulations with $N=128\mathrm{k}$, the fluctuation spectra (i.e. the Fourier transform amplitude) of both $\eta_{\mathrm{v}}(t_0)$ and $\beta_{\mathrm{v}}(t_0)$ are approximately a power law with index of $-1.25$. The error in each of the coefficients $(\alpha, \eta_{\rm s},\eta_{\rm v},\beta_{\rm v})$ has three components: (i) statistical error due to the finite number of particles in the region considered, (ii) error due to the fitting procedure, and (iii) systematic error due to the fact that the coefficients really do fluctuate in time and differ between different realizations, as discussed above. The first can be estimated by splitting the particles in the region arbitrarily into two or more groups, and obtain the fit for each group separately. The second components can be estimated from the least squares procedure. Finally, the systematic error can be estimated from the width of the distribution of the coefficients obtained at different values of $t_0$ and for different realizations. We found that the first two components of the error are much smaller than the third, and will thus be ignored. The error values presented in Table \ref{tab:results-N}
should thus be understood as systematic variations in the 2-body relaxation and VRR processes. We note that the relative error in $\eta_{\mathrm{v}}$ is about a factor of 2 larger than the other non-resonant coefficients $(\alpha, \eta_{\rm s})$.

Table \ref{tab:results-N} also shows the upper limits on the scalar resonant relaxation coefficient $\beta_{\mathrm{s}}$ and the equivalent parameter for the coherent change in energy, $\tilde{\alpha}$. These parameters are obtained by fitting $\delta E(\tau)$ and $\delta L_{\mathrm{s}}(\tau)$ with the following expression, analogous to Equation~(\ref{eq:rms-deltaLv})
\begin{align}
\mathrm{rms}(\delta E) &= \frac{m}{M}\sqrt{N}(\alpha\sqrt{\tau} + \tilde{\alpha}\tau)\\
\mathrm{rms}(\delta L_\mathrm{s}) &= \frac{m}{M}\sqrt{N}(\eta_\mathrm{s}\sqrt{\tau} + \beta_\mathrm{s}\tau)
\end{align}
As expected, the values of $\tilde{\alpha}$ and $\beta_\mathrm{s}$ are scattered around zero and are very small in magnitude compared to the other coefficients. The upper limits in Table \ref{tab:results-N}
are one standard deviation. Table \ref{tab:results-N} shows that the different simulations agree with each other (see discussion on the lack of $N$-dependence in Section \ref{sec:discussion}).

\begin{table*}
\begin{center}
\begin{tabular}{|c|c|c|c|c||c|c|}
\hline  & $\alpha$ & $\eta_{\mathrm{s}}$ & $\eta_{\mathrm{v}}$ & $\beta_{\mathrm{v}}$ & $\alpha_{\mathrm{RR}}$ & $\beta_{\mathrm{s}}$\\
\hline \hline 128k\_1 & $7.36\pm0.24$ & $8.60\pm0.26$ & $11.6\pm0.66$ & $1.18\pm0.43$ & $<0.06$ & $<0.06$\\
\hline 128k\_2 & $7.41\pm0.30$ & $8.59\pm0.17$ & $11.7\pm0.69$ & $1.09\pm0.46$ & $<0.07$ & $<0.06$\\
\hline 256k & $7.30\pm0.16$ & $8.62\pm0.19$ & $11.3\pm0.57$ & $1.42\pm0.38$ & $<0.04$ & $<0.04$\\
\hline 512k & $7.32\pm0.15$ & $8.82\pm0.23$ & $11.5\pm0.94$ & $1.33\pm0.50$ & $<0.04$ & $<0.04$\\
\hline \hline Combined & $7.35\pm0.22$ & $8.67\pm0.22$ & $11.5\pm0.73$ & $1.26\pm0.44$ &  & \\
\hline
\end{tabular}
\end{center}
\caption{Measured dimensionless coefficients (Equations. \ref{eq:rms-deltaE}-- \ref{eq:rms-deltaLv}) from simulations for Region II. The two columns on the right are upper limits to the resonant coefficient for the energy and angular momentum magnitude as explained in the text. The errors represent the width of the distribution of each coefficient, and not the fitting error, which is typically much smaller. The combined result in the bottom row is a simple unweighted average of the above rows; no $N$ dependence is observed.\label{tab:results-N}}
\end{table*}

Table \ref{tab:results-region} shows the same coefficients but at all four regions specified in Table \ref{tab:regions}, and only for the largest simulation with $N=512\mathrm{k}$. The values differ by up to about an order of magnitude among the different regions. It seems empirically that $\beta_{\mathrm{v}}/\eta_{\mathrm{v}}\approx0.1$ in all regions. For comparison, the ratio $\beta_{\mathrm{v}}/\eta_{\mathrm{v}}$ is $\approx0.9$ \citep{Rauch+96} and $\approx1.1$ \citep{Eilon+09}
for the Keplerian potential case; for the isochrone potential case (which is similar to a Plummer potential in that it has a flat core rather than a singularity) \citeauthor{Rauch+96} get $\beta_{\mathrm{v}}/\eta_{\mathrm{v}}\approx0.2$. The fluctuations in $\beta_{\mathrm{v}}$, expressed by the standard deviation, are consistently $\approx40\%$ in our simulations.

\begin{figure*}
\includegraphics{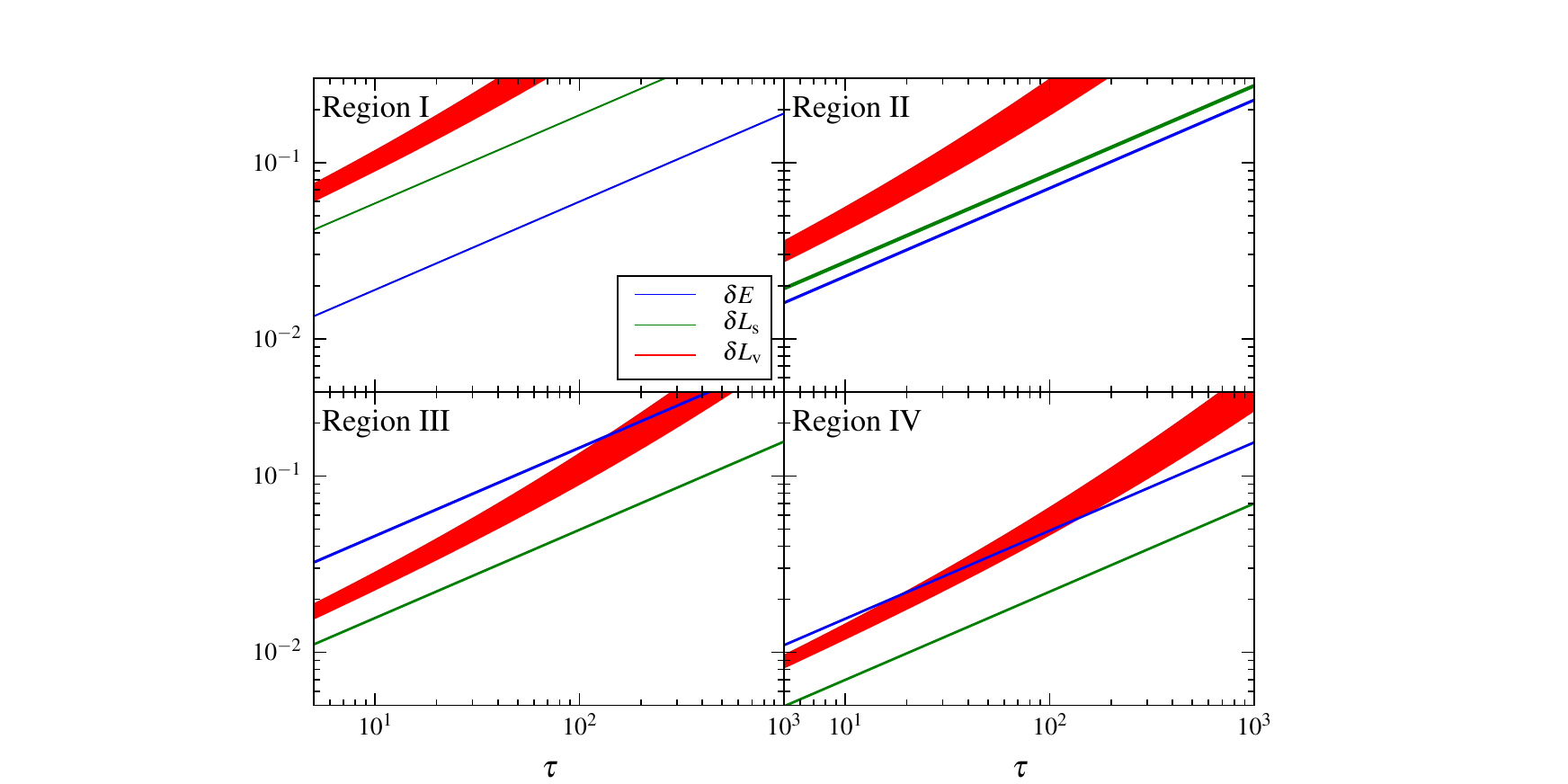}
\caption{Similar to Figure~\ref{fig:delta-example} for all four regions, but instead of showing the quantities measured from the data, plots of Equations (\ref{eq:rms-deltaE})--(\ref{eq:rms-deltaLv}) are shown, where the values for each of the parameters is taken from Table \ref{tab:results-region} and $N=1024\mathrm{k}$. The lines appear thick because of the range of each parameter, as given be its standards error in the Table~\ref{tab:results-N}.\label{fig:deltas-regions}. We note that the reorientation angle of angular momentum vector directions is $(L^2/L_c^2)\delta L_{\rm v}$, which is 0.4, 0.5, 0.06 and 1.0 times $\delta L_{\rm v}$ for the four regions in order respectively.   }
\end{figure*}

Figure~\ref{fig:deltas-regions} is similar to Figure~\ref{fig:delta-example} for all four regions, but instead of showing the quantities measured from the simulated data, it plots of Equations (\ref{eq:rms-deltaE})--(\ref{eq:rms-deltaLv}) are shown, where the values for each of the parameters is taken from their fitting values shown in Table \ref{tab:results-region} and $N=1024\mathrm{k}$. The lines appear thick because of the allowed range of each parameter, as given be their standard error in the Table.

\begin{table*}
\begin{center}
\begin{tabular}{|c|c|c|c|c|c|}
\hline  & $\alpha$ & $\eta_{\mathrm{s}}$ & $\eta_{\mathrm{v}}$ & $\beta_{\mathrm{v}}$ & $N_\mathrm{crit}$\\
\hline \hline I & $6.14\pm0.08$ & $19.0\pm0.24$ & $25.5\pm1.59$ & $2.52\pm0.98$ & $6.7\times10^4$\\
\hline II & $7.32\pm0.15$ & $8.82\pm0.23$ & $11.5\pm0.94$ & $1.33\pm0.50$ & $9.9\times10^3$\\
\hline III & $14.8\pm0.29$ & $5.07\pm0.09$ & $6.79\pm0.38$ & $0.47\pm0.20$ & $9.6\times10^3$\\
\hline IV & $5.02\pm0.09$ & $2.26\pm0.04$ & $3.58\pm0.17$ & $0.22\pm0.09$ & $3.4\times10^3$\\
\hline
\end{tabular}
\end{center}
\caption{ Measured dimensionless coefficients as in Table~\ref{tab:results-N} but for different regions in the $N=512\mathrm{k}$ simulation. $N_\mathrm{crit}$ is the number of particles beyond which VRR becomes dominant as discussed in Section \ref{sec:Ncrit}.\label{tab:results-region}}
\end{table*}

In the limit that the directions of angular momentum vectors $\bm{L}_i$ are reoriented during VRR while $E_i$ and $L_i$ change much more slowly, VRR may be represented as mixing on a spherical surface of the unit-normalized angular momentum vectors, $\bm{L}_i/L_i$. While Equation~(\ref{eq:rms-deltaLv}) breaks down when the orbital planes have on average changed by $\sim 1$ radian, a better description is provided by the angular correlation function $C(\mu,\Delta t)$ which specifies how the angular momentum direction distribution function at time $t_0$ correlates with that at time $t_0+\Delta t$ for angular separations $\mu=\cos\theta$ \citep{Kocsis+15}. In Appendix~\ref{app:correlation} we show that
\begin{equation}
C({\mu,\Delta t}) = \sum_{\ell=0}^{\infty} C_{\ell}(\Delta t) P_{\ell}(\mu)
\end{equation}
where $P_{\ell}(\mu)$ are Legendre polynomials, and the coefficients depend on the multipole moments of the distribution $C_{\ell}(\Delta t)=\sum_{m=-{\ell}}^{\ell}\langle Y_{\ell m}(t_0) \rangle\,\langle Y_{\ell m}(t_0+\Delta t) \rangle $, where $\langle Y_{\ell m}(t)\rangle$ are spherical harmonics averaged over the distribution of angular momentum directions at a given time $t$, and we average over $t_0$. \citet{Kocsis+15} have shown that if the angular momentum vectors exhibit independent Brownian motion on the sphere with diffusion coefficient $D$, then
\begin{equation}\label{eq:cellVell}
C_{\ell}(\Delta t)=\frac{2\ell+1}{4\pi} e^{-\frac14 \ell(\ell+1)V_{\ell}}
\end{equation}
where $V_{\ell}=D\Delta t$ for all $\ell$. More generally, for an arbitrary random process, the correlation function $C_{\ell}$ may still be represented with Equation~\eqref{eq:cellVell} using $V_{\ell}(\Delta t)$, which may be unequal and nonlinear. In the coherent phase of VRR, $V_{\ell}(\Delta t)\propto \Delta t^2$, while for incoherent VRR $V_{\ell}(\Delta t) \propto \Delta t$ until the distribution function becomes fully uncorrelated for the given number of particles \citep{Kocsis+15}. The slope of a quadratic time dependence is analogous to $(L_\mathrm{c}^2/L^2)N^{-1}\beta_{Lv}^2$ in equation~\eqref{eq:rms-deltaLv}, and the slope of the following linear time dependence specifies the rate of incoherent mixing during VRR, which was neglected in equation~\eqref{eq:rms-deltaLv}. We refer the readers to Appendix~\ref{app:correlation} for a details.

\begin{figure*}
\includegraphics{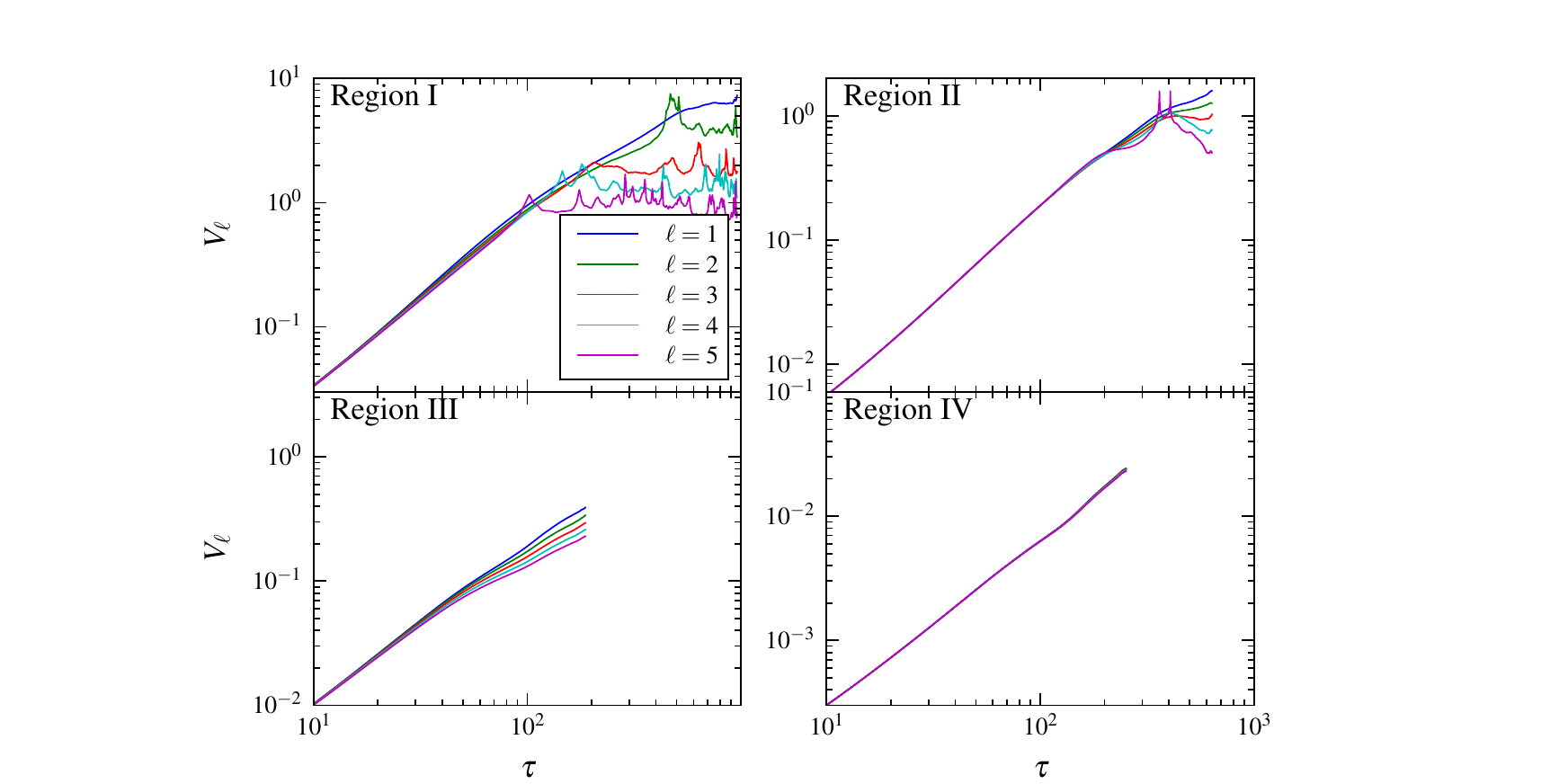}
\caption{The five first moments of the angular variance $V_{\ell}(\tau)$ for each of the four regions in $(E,L)$ space, measured in our $N=512\mathrm{k}$ simulation.\label{fig:angular-correlation}}
\end{figure*}

The four panels of Figure~\ref{fig:angular-correlation} show the angular variance $V_{\ell}(\tau)$ for the four regions measured in our $N=512\mathrm{k}$ simulation. The result is qualitatively very similar to figures 8 and 11 in \citet{Kocsis+15} which describe galactic nuclei, in that $V_{\ell}$ is described by an approximately quadratic dependence at early times, which changes to a shallower roughly linear dependence at later times until saturation sets in. The values of $V_{\ell}$ are similar for different $\ell$. Note however, that unlike \citet{Kocsis+15}, our simulation resolves 2-body relaxation as well, which manifests as a linear $V_{\ell}$ before coherent VRR sets in.

\section{Non-resonant relaxation}\label{sec:2body}

The incoherent diffusion coefficients $\alpha$, $\eta_{\mathrm{s}}$, and $\eta_{\mathrm{v}}$ are trivially related to the orbitally-averaged diffusion coefficients $D_{E^{2}}$, $D_{L_{\mathrm{s}}^{2}}$, and $D_{L_{\mathrm{v}}^{2}}$ as follows
\begin{equation}\label{eq:alpha}
\alpha=\frac{\sqrt{{D_{E^{2}}NP  \vphantom{D_{L_{\mathrm{s}}^{2}}}   }}}{|E|};\ \eta_{\mathrm{s}}=\frac{\sqrt{{D_{L_{\mathrm{s}}^{2}}NP}}}{L_\mathrm{c}};\ \eta_{\mathrm{v}}=\frac{\sqrt{{D_{L_{\mathrm{v}}^{2}}NP}}}{L_\mathrm{c}}.
\end{equation}
In Paper I, we calculated $D_{E^{2}}$ and $D_{L_{\mathrm{s}}^{2}}$ from the Chandrasekhar scattering theory for different values of $E$ and $L$ in a Plummer sphere. The diffusion of the angular momentum direction, $D_{L_{\mathrm{v}}^{2}}$ can be calculated in much the same way, and this calculation is presented in Appendix \ref{sec:appendix-a}.

In order to compare the Chandrasekhar's scattering theoretical results with the measured coefficients, we need to take into account the fact that the regions we define in $(E,L)$-space are not very narrow around their central values (unlike in Paper I). This means that in order for the comparison to be meaningful, the orbitally-averaged diffusion coefficients need to be averaged in a weighted way within each region. The results of this calculation are shown in Table \ref{tab:2body-theory}. The theory, in principle, has no free parameters. However, the Coulomb logarithm folds much of the uncertainly and in practice is used as a fudge factor. By dividing $\eta_{\mathrm{s}}$ and $\eta_{\mathrm{v}}$ by $\alpha$, we cancel the theoretical dependence on the Coulomb logarithm and are able to compare the simulation with the theory.

In addition to the $N$-body simulation, we performed Monte Carlo simulations, where pairs of stars do not interact directly but only in a statistical way. In Paper I we also use this method to follow the long-term time evolution of the probability amplitude for a particle to transition from one state to another (the propagator). There, we used the Hénon version of the Monte Carlo method developed originally by \citealt{Henon71}, improved by \citealt{Stodolkiewicz86}; this is the most commonly used version. Here, articles in adjacent radii are randomly chosen to interact at timesteps equal to a fraction of the 2-body relaxation time, and the relative velocity change is amplified to statistically mimic the effect of a large number of scattering events. A core assumption of this method is that the cluster is spherically symmetric, and the MOCCA implementation of this method \citep{Giersz+13}, used in Paper I, particles are not assigned full 6D phase space coordinates, but rather just radius, radial and tangential velocities. Thus, it cannot be used to study relaxation of the orbital planes.

Another version of the Monte Carlo method is due to \citet{Spitzer+71}, often called the Spitzer or Princeton version. The significant difference to the Hénon version is that here, the timestep is a fraction of the dynamical time, rather than the 2-body relaxation time, and perturbations to each star's velocity are applied independently by assuming a Maxwellian distribution of background stars. Since in this method all position and velocity components of the particles are known (much like in an $N$-body simulation), the orientation of the orbital planes could be followed.

We used the Monte Carlo code RAGA \citep{Vasiliev15} that implements the Spitzer version of the method and performed an $N=512\mathrm{k}$ simulation with identical initial conditions to that of the $N$-body simulation. We analyzed it in exactly the same way, by considering different regions in $(E,L)$ space and calculating the rms of $\delta E$, $\delta L_\mathrm{s}$, and $\delta L_\mathrm{v}$ as functions of time, fitting for $\alpha$, $\eta_\mathrm{s}$, $\eta_\mathrm{v}$, and $\beta_\mathrm{v}$. Unlike other Monte Carlo codes, RAGA calculates the potential through the self-consistent field method \citep{Clutton-Brock73,Hernquist+92,Meiron+14} that relaxes the assumption of spherical symmetry to a degree by calculating the gravitational field through a multipole expansion that is cut at some $\ell_\mathrm{max}$. The results presented in Table \ref{tab:2body-theory} are for $\ell_\mathrm{max}=0$ (only the monopole, so spherical symmetry is assumed), but we also tested $\ell_\mathrm{max}=2$ (quadrupole). The results were similar in both cases, with specifically $\beta_\mathrm{v}$ consistent with zero. This is expected for the monopole case, as there are no torques in the system, but somewhat surprising for the quadrupole case.

While the $N$-body simulations and RAGA agree quite well when it comes to the non-resonant relaxation of different integrals of motion in different regions in $(E,L)$ space, there is some visible tension with the values calculated from Chandrasekhar's theory.

\begin{table*}
\begin{center}
\begin{tabular}{|c|c|c||c|c||c|c|}
\hline  & \multicolumn{2}{c||}{$N$-body} & \multicolumn{2}{c||}{Theory} & \multicolumn{2}{c|}{RAGA}\\
\hline  & $\eta_{\mathrm{s}}/\alpha$ & $\eta_{\mathrm{v}}/\alpha$ & $\eta_{\mathrm{s}}/\alpha$ & $\eta_{\mathrm{v}}/\alpha$ & $\eta_{\mathrm{s}}/\alpha$ & $\eta_{\mathrm{v}}/\alpha$\\
\hline \hline I & $3.10\pm0.06$ & $4.15\pm0.26$ & $2.85$ & $3.72$ & $2.79\pm0.05$ & $4.27\pm0.07$\\
\hline II & $1.20\pm0.04$ & $1.57\pm0.13$ & $1.14$ & $1.60$ & $1.14\pm0.02$ & $1.76\pm0.03$\\
\hline III & $0.34\pm0.01$ & $0.46\pm0.03$ & $0.23$ & $0.26$ & $0.31\pm0.01$ & $0.45\pm0.02$\\
\hline IV & $0.45\pm0.01$ & $0.71\pm0.04$ & $0.53$ & $0.90$ & $0.44\pm0.01$ & $0.78\pm0.02$\\
\hline
\end{tabular}
\end{center}
\caption{The ratios of the non-resonant coefficients $\eta_{\mathrm{s}}$ and $\eta_{\mathrm{v}}$ with $\alpha$ for different regions and different methods.\label{tab:2body-theory}}
\end{table*}

\section{Discussion}\label{sec:discussion}

We discuss the implications of the simulations to real globular clusters as follows. First we make inferences on the region of parameters where VRR operates in the simulations, then we discuss extrapolations to real systems, and compare with the typical parameters of observed globular clusters. Then we discuss whether this effect may operate in dwarf galaxies and finally discuss the possibility of black hole disks in globular clusters.

\subsection{Constraints on VRR in globular clusters}\label{sec:Ncrit}

According to Equation~(\ref{eq:rms-deltaLv}), the reorientation of angular momentum vectors is dominated by coherent torques on timescales when $\tau \geq (\eta_\mathrm{v}/\beta_\mathrm{v})^2$. As also noted in Section~\ref{sec:simulations}, the expression for the growth of $\mathrm{rms}(\delta L_\mathrm{v})$ is only an appropriate model when the quantity is much smaller than unity. We define three timescales through the extrapolation of Equations~(\ref{eq:rms-deltaE})--(\ref{eq:rms-deltaLv}) to unity and substituting $M=Nm$, these are
\begin{align}\label{eq:trxE}
t_{\mathrm{rx},E} & = N P/\alpha^{2}\\
t_{\mathrm{rx},L_{\mathrm{s}}} & =NP/\eta_{\mathrm{s}}^{2}\label{eq:trxLs}\\
t_{\mathrm{rx},L_{\mathrm{v}}} & =\frac{1}{4\beta_{\mathrm{v}}^{2}}\left(-\eta_{\mathrm{v}}+\sqrt{\eta_{\mathrm{v}}^{2}+4\beta_{\mathrm{v}}\sqrt{N}}\right)^{2}P\label{eq:trxLv}
\end{align}
We note that $t_{\mathrm{rx},E}$ in equation~\eqref{eq:trxE} specifies the time when $\mathrm{rms}[(E-E_0)/E_0]=1$, which is different from the definition in Paper I, where the normalization was with respect to the average kinetic energy in the region. Similarly, note that the relaxation time of angular momentum is often defined in the literature as $(L/L_\mathrm{c})^2 t_{\mathrm{rx},L_{\mathrm{s}}}$ and similarly for $t_{\mathrm{rx},L_{\mathrm{v}}}$ with respect to the adopted definition \citep{2012EPJWC..3905001A,Kocsis+15}. Indeed, the timescale for the angular momentum vector to reorient by an angle of order 1 radian is $(L/L_\mathrm{c})^2 t_{\mathrm{rx},L_{\mathrm{v}}}$, which is significantly shorter than $t_{\mathrm{rx},L_{\mathrm{v}}}$ given by equation~(\ref{eq:trxLv}) for lower angular momentum orbits; e.g., by a factor 16.4 for Region III.

Equation~\eqref{eq:trxLv} is more complicated than equations~\eqref{eq:trxE}--\eqref{eq:trxLs} because it is a solution to a quadratic equation. It has two asymptotic expressions
\begin{equation}\label{eq:trxLv1}
t_{\mathrm{rx},L_{\mathrm{v}}}=\begin{cases}
NP/\eta_{\mathrm{v}}^{2} & N\ll \eta_{\rm v}^4/\beta_{\rm v}^2\\
\sqrt{N}P/\beta_{\mathrm{v}} & N\gg \eta_{\rm v}^4/\beta_{\rm v}^2
\end{cases}
\end{equation}

Rapid reorientation happens at fixed energy and angular momentum if
\begin{equation}\label{eq:tcoh2}
\left(\frac{\eta_{\rm v}}{\beta_{\rm v}}\right)^2 P \lesssim t \lesssim \min( t_{\mathrm{rx},E} , t_{\mathrm{rx},L_{\mathrm{s}}}, t_{\mathrm{rx},L_{\mathrm{v}}}  ).
\end{equation}
VRR operates if coherent torques dominate the reorientation in globular clusters if the interval $(\eta_\mathrm{v}/\beta_\mathrm{v})^2 P \lesssim t \lesssim t_{\mathrm{rx},L_{\mathrm{v}}}$ is non-empty (i.e., there exist values of $\delta L_\mathrm{v}$ corresponding to it). This is satisfied when
\begin{equation}
N \gtrsim N_\mathrm{crit}\equiv \frac{\eta_{\rm v}^4}{\beta_{\rm v}^2}.
\end{equation}
The reorientation of orbital planes may happen at a nearly fixed energy and angular momentum magnitude, so that the orbits rotate as rigid bodies if
\begin{equation}\label{eq:tcrit}
\frac{L^2}{L_\mathrm{c}^2}t_{\mathrm{rx},L_\mathrm{v}}\ll \mathrm{max}(t_{\mathrm{rx},E},t_{\mathrm{rx},L_\mathrm{s}})
\end{equation}
which happens if\footnote{We find that $(L_\mathrm{c}/L)\eta_{\rm v}<\max(\alpha,\eta_{\rm s})$ for all regions considered, so the low $N$ asymptote of Equation~\eqref{eq:trxLv1} can never satisfy Equation~\eqref{eq:tcrit}.}
\begin{equation}
N \gg N_\mathrm{crit,\,rigid} \equiv \max\left( \frac{\eta_{\rm v}^4}{\beta_{\rm v}^2}, \frac{L^4\eta_{\rm s}^4}{L_\mathrm{c}^4\beta_{\rm v}^2}, \frac{L^4\alpha^4}{L_\mathrm{c}^4\beta_{\rm v}^2}, \right)
\end{equation}
Using Table~\ref{tab:results-region} for the parameters in Region II (recall that this is the region around the geometrical median of $E$ and $L$ values of a Plummer sphere), $N_\mathrm{crit,\,rigid} = N_\mathrm{crit} \approx 1.1\times 10^4$.\footnote{This is derived from the average of all simulations as appearing in Table~\ref{tab:results-N} rather than the values for Region II in the $N=512\mathrm{k}$ simulations appearing in the second row of Table~\ref{tab:results-region}.}

\subsection{Correcting for softening}\label{sec:Coloumb}
The dimensionless parameters $\alpha$, $\eta_{\mathrm{s}}$, and $\eta_{\mathrm{v}}$ are related to 2-body relaxation, they are proportional to the square root of the diffusion coefficients (Equation~\ref{eq:alpha}), which are in turn proportional to the Coulomb logarithm. However, in Section \ref{sec:simulations} we have shown that the dimensionless parameters $\alpha$, $\eta_{\mathrm{s}}$, and $\eta_{\mathrm{v}}$ describing the incoherent diffusion of $E$, $L$, and $\bm{L}$ are independent of $N$ (see Table~\ref{tab:results-N}). This is not surprising because the simulations used softened 2-body interactions, with softening lengths of $\epsilon=3\times10^{-4}$, which is much larger than the $90^{\circ}$ deflection impact parameter $b_{90}$, which is of the order of $GM/(3N\sigma^{2})$ where $\sigma$ is the local (one dimensional) velocity dispersion (in a Plummer model within the half-mass radius, it is $\approx 0.47$). Therefore $b_{90}$ would be at most $\sim10^{-5}$ in the simulations presented here (all numbers in Hénon units, see footnote \ref{fn:units}). Thus, the Coulomb logarithm $\ln\Lambda=\ln(b_{\mathrm{max}}/b_{\mathrm{min}})$ is constant in the models we explored, and cannot be distinguished from the other multiplicative constants. We assume that $b_{\mathrm{max}}=r_{0}$ and that $b_{\mathrm{min}}=\epsilon$, which gives $\ln\Lambda\approx7.58$. To scale our results to real globular clusters, we first have to correct for the missing $N$ dependence:
\begin{equation}
X_{\mathrm{real}}= \left(\frac{\ln\Lambda_{\mathrm{real}}}{\ln\Lambda_{\mathrm{sim}}}\right)^{1/2}X = \left[\frac{\ln (0.4 N_{\rm real})}{\ln (r_{0}/\epsilon)}\right]^{1/2}X
\end{equation}
where $X$ denotes $\alpha$, $\eta_{\rm s}$, or $\eta_{\rm v}$, and the subscript ``real'' indicates that the quantity is corrected. The Coulomb logarithm for real clusters $\ln\Lambda_{\mathrm{real}}$ needs to be evaluated by other means, we adopt the common practice \citep{Spitzer87} that $\ln\Lambda_{\mathrm{real}}=\ln(0.4N_{\mathrm{real}})$ where $N_{\rm real}$ denotes the number of stars in the cluster. The dimensionless coefficient $\beta_{\mathrm{v}}$ is not expected to be affected by this modification since it is set by global torques for low order multipole moments. We assume that it does not need to be corrected for the $N$ dependence.

\subsection{Relaxation in observed globular clusters}
\begin{figure}
\includegraphics[width=1\columnwidth]{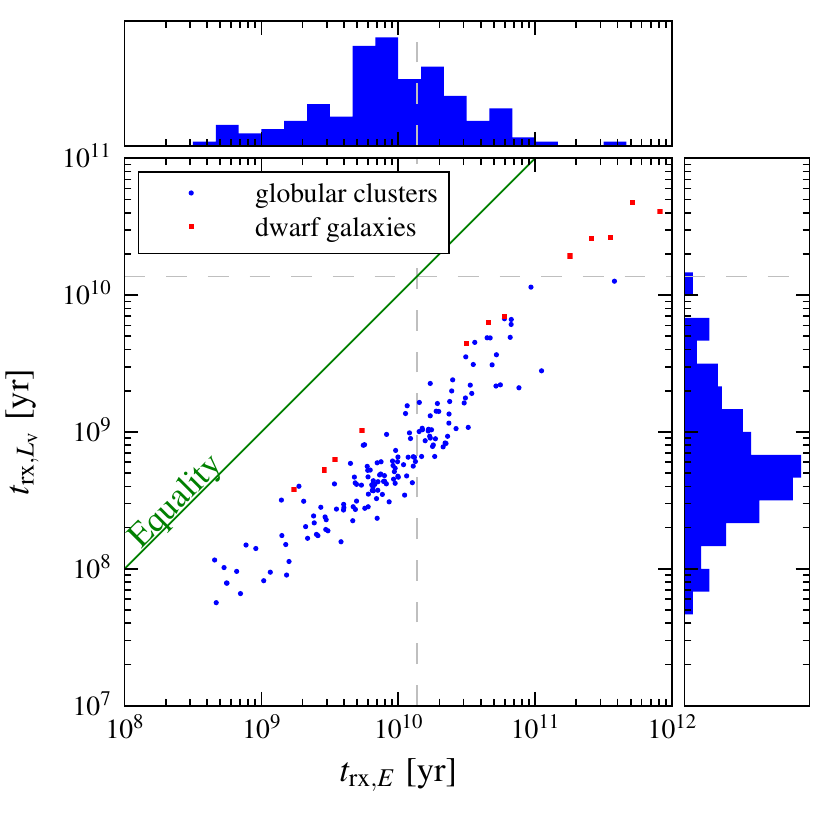}
\caption{The energy relaxation time $t_{\mathrm{rx},E}$ and angular momentum direction relaxation time $t_{\mathrm{rx},L_\mathrm{v}}$ for Region II stars (most typical) in globular clusters (blue circles) and Local Group dwarf galaxies (red squares). The histograms show the globular cluster only counts in logarithmic time bins. The two dashed gray lines show the Hubble time.\label{fig:real-data}}
\end{figure}

In the \citet[][2010 edition]{Harris96} catalog of galactic globular clusters, the relaxation time is estimated for each object (152 objects have valid values for the relaxation time) in the following way (based on \citealt{Djorgovski93})
\begin{align}
t_{\mathrm{rx}}=&(2.055\times10^{6}\,\mathrm{yr})\,[\ln(0.4N)]^{-1}\left(\frac{\langle m\rangle}{\mathrm{M}_{\odot}}\right)^{-1}\left(\frac{M}{\mathrm{M}_{\odot}}\right)^{1/2}\nonumber\\&\times \left(\frac{r_{\mathrm{h}}}{1\,\mathrm{pc}}\right)^{3/2}\label{eq:trx-harris}
\end{align}
where the following assumptions are made:
\begin{enumerate}
\item The mass-to-light ratio is $\Upsilon=2\,\mathrm{M}_{\odot}\mathrm{L}_{\odot}^{-1}$; the cluster's luminosity is calculated from the absolute magnitude column. \item The average stellar mass is $\langle m\rangle=(1/3)\,\mathrm{M}_{\odot}$. \item The half-mass radius $r_{\mathrm{h}}$ is calculable from the half-light radius and the distance information.
\end{enumerate}

In Section \ref{sec:Ncrit} we have derived the minimum number of stars where VRR may be expected to operate. When correcting for the $N$ dependence of $\eta_{\mathrm{v}}$ as explained in Sec.~\ref{sec:Coloumb} above, we get a non-linear algebraic equation in $N$, the solution of which is $N_{\mathrm{crit,real}}\approx1.44\times10^{4}$ for the values corresponding to Region II (the most typical orbits). Table~\ref{tab:results-region} shows $N_{\mathrm{crit}}$ for the four regions in the 512k simulation. Relative to two-body relaxation, VRR is least efficient in the core of the cluster (region I), but even there $N_{\mathrm{crit}}<10^5$.

\textit{We conclude that clusters with $N>1.44\times 10^{4}$ are affected by VRR.}
This number is significantly smaller than the median for the globular clusters in the Harris catalog (given the above assumptions on the mean stellar mass and mass-to-light ratio), which is $\sim3\times10^{5}$. Thus, most globular clusters are strongly subjected to VRR. In simulations, it is possible to artificially lower this threshold by increasing the softening length, use a mean-field method to calculate the gravitational interactions such as a Self-Consistent Field method, or use smooth background potential instead accounting of full ``$N^2$'' interactions. In all these cases 2-body relaxation is suppressed, while VRR is not.

Finally, when considering the corrections for $\alpha$ and $\eta_{\mathrm{s}}$, our expressions for $t_{\mathrm{rx},E}$ and $t_{\mathrm{rx},L_{\mathrm{s}}}$ are identical in form to Equation~(\ref{eq:trx-harris}), but with different coefficients (depending on whether the energy or angular momentum magnitude is considered). For the 152 GCs in the Harris catalog, our result is that $t_{\mathrm{rx},E}$ for Region II orbits is $9.0\times$ larger than the values given in the Harris catalog, and $t_{\mathrm{rx},L_{\mathrm{s}}}$ is $6.5\times$ larger.

Figure \ref{fig:real-data} shows the angular momentum direction relaxation time $t_{\mathrm{rx},L_{\mathrm{v}}}$ versus the energy relaxation time $t_{\mathrm{rx},E}$ for the 152 objects in the Harris catalog (blue circles) and 12 objects from the \citet{McConnachie12} catalog of dwarf galaxies in and around the Local Group (red squares). The timescale is calculated under the assumptions listed above, using the dimensionless coefficients corresponding to Region I in our simulation, where the orbital period is $P=5.88\sqrt{R^3/(GM)}$ (the relaxation times are proportional to $P$). For the globular clusters, we find that $t_{\mathrm{rx},E}$ is larger than $t_{\mathrm{rx},L_{\mathrm{v}}}$ by a factor of between 4 and 40 (the median is $\sim15$). While approximately one third of the objects have $t_{\mathrm{rx},E}>t_{\mathrm{H}}$, no objects have $t_{\mathrm{rx},E}$ longer than $t_{\mathrm{H}}$.

\subsection{Application to dwarf galaxies}
Spherical galaxies are subjected to the same physical processes. While at their very center, the supermassive black hole, when present, dominates the dynamics, its influence does not extent to beyond a small fraction of the galaxy's size. The energy relaxation time of galaxies may exceed the Hubble time, $t_\mathrm{H}=13.7\times10^9$\,yr, by orders of magnitude, but this is not the case for many dwarf spheroidal galaxies and often for spherical components of spiral galaxies, such as the Milky Way.

An important difference between dwarf spheroidal galaxies and globular clusters is that the former contain a dark gravitational component which may exceed the stellar gravity by a large factor. How this affects the relaxation time depends on the nature of this dark component. If it is made of subatomic or microscopic particles in a spherical geometry, then it does not affect either diffusion coefficients or global torques, but only the dynamical time (i.e. only $P$ is changed in the definition of $\tau$ in Equations~\ref{eq:rms-deltaE}--\ref{eq:rms-deltaLv}). If instead it is made of stellar-mass particles, then it participates in both 2-body and resonant relaxation processes, increasing their timescale by making $N$ effectively much larger. If Modified Newtonian Gravity is at work, then it is not easy to predict how both relaxation processes would be affected.

Another important caveat to remember is that our simulations are based on the Plummer model, which is just an approximation of a globular cluster or a dwarf spheroidal galaxy; in reality, different clusters will have different mass profiles. We also assumed a single mass species, no binaries and importantly, perfect spherical symmetry, where the global torques only come from the $\sqrt{N}$ noise rather than the ellipsoidal or triaxial shape of the cluster, but a cluster's asphericity may strongly affect the angular momentum relaxation times. The dwarf spheroidal galaxies in Figure \ref{fig:real-data} were not filtered by their ellipticities. For the 12 galaxies shown, 10 had ellipticity values in the catalog, the median of which was 0.48.

We calculated the relaxation times for dwarf galaxies under the same assumptions, but evaluated $N$ from the stellar mass and used the dynamical mass to calculate the scaling of the time unit. In Figure~\ref{fig:real-data} we show that some dwarf galaxies in \citet{McConnachie12} have VRR times significantly smaller than the Hubble time. In these dwarf galaxies, VRR is expected to operate as in globular clusters well outside of the radius of influence of the central massive black hole.

\subsection{Implications of VRR in globular clusters}
Our results suggest that the half-mass relaxation times may be longer than previously thought. Combined with the result from Paper I that showed that mixing timescale (of the energy and angular momentum magnitude) is approximately 10 times longer than the energy half-mass relaxation time, many old globular clusters could still retain memory of their initial conditions.

VRR has been widely discussed previously for nuclear star clusters. Recently, \citet{Szolgyen+18} examined the case of a star cluster around a supermassive black hole forming by 16 episodes of star formation or globular cluster infall using a Monte Carlo Markov Chain simulation of VRR which neglected 2-body relaxation by construction. They showed that in the statistical equilibrium configuration of such a system, massive stars and stellar mass black holes form a warped disk, while low mass stars are spherically distributed. This anisotropic mass segregation is driven by coherent mutual gravitational torques in the system, quantified in this paper by $\beta_{\mathrm{v}}$. This process may be quenched by 2-body encounters, quantified here by $\eta_{\mathrm{v}}$.

If globular clusters form in a similar way, in the sense that infalling material accumulates in a few discrete episodes, a disk of heavy object will form in globular clusters. Indeed, we found that VRR operates in globular clusters if its mass exceeds $10^4\,{\rm M}_{\odot}$. The VRR timescale in globular clusters is much shorter than the 2-body relaxation time by a factor 4--40 (see Figure~\ref{fig:real-data}), and it is less than a Hubble time for all globular clusters. We are currently running multimass $N$-body simulations to see if a disk of black holes is indeed maintained in VRR.

We conclude that VRR operates efficiently in most globular clusters and in some low mass dwarf galaxies.

\acknowledgments This work has been supported by the European Research Council under the European Union's Horizon 2020 Programme, ERC-2014-STG grant GalNUC 638435. The special GPU accelerated supercomputer Laohu at the Centre of Information and Computing at National Astronomical Observatories, Chinese Academy of Sciences, funded by Ministry of Finance of the People's Republic of China under the grant ZDYZ2008-2, has been used for the largest simulations. Some calculations were carried out on the NIIF HPC cluster at the University of Debrecen, Hungary.

\bibliographystyle{yahapj}
\bibliography{main}

\appendix

\section{Diffusion of angular momentum direction}\label{sec:appendix-a}

First we mirror Appendix A of Paper I, but instead of the the local diffusion of angular momentum \emph{magnitude}, we calculate that of the \emph{direction}. We do this by writing the mean square change in the angular momentum vector during a short encounter as a function of the mean square velocity changes parallel and perpendicular to the original velocity direction. In other words, express $\langle|\Delta\bm{L}|^{2}\rangle$ as a function of $\langle(\Delta v_{\parallel})^{2}\rangle$ and $\langle(\Delta v_{\bot})^{2}\rangle$. Since we are only computing the average change during a single short encounter (``local diffusion'') our expressions will depend on phase space coordinates (namely $r$, $v$ and $v_{r}$). In the next step we will integrate over them to get the orbital averaged coefficients. Also note that we are interested in the square magnitude of the difference vector, not the square change in the vector's magnitude (as in Paper I), thus $|\Delta\bm{L}|^{2}\equiv|\bm{L}_{2}-\bm{L}_{1}|^{2}$.

We start by writing the angular momentum vector before the encounter $\bm{L}_{1}=\bm{r}\times\bm{v}_{1}$. Since the encounter occurs over a very short period, $\bm{r}$ does not change and therefore does not need to be subscripted. After the encounter, the angular momentum vector is $\bm{L}_{2}=\bm{r}\times\bm{v}_{2}$ with
\begin{equation}
\bm{v}_{2}=\bm{v}_{1}+(\Delta v_{\parallel})\hat{\bm{v}}_{1}+(\Delta v_{\bot})\hat{\bm{u}}_{1}
\end{equation}
where $\hat{\bm{v}}_{1}=\bm{v}_{1}/v_{1}$ a unit vector in the direction of $\hat{\bm{v}}_{1}$, and $\hat{\bm{u}}_{1}$ an unknown unit vector perpendicular to it. The new angular momentum following some simple algebra is
\begin{equation}
\bm{L}_{2}=\left[1+\frac{(\Delta v_{\parallel})}{v_{1}}\right]\bm{L}_{1}+(\Delta v_{\bot})(\bm{r}\times\hat{\bm{u}}_{1})
\end{equation}
the difference vector is
\begin{equation}
\Delta\bm{L}=\bm{L}_{2}-\bm{L}_{1}=\frac{\Delta v_{\parallel}}{v_{1}}\bm{L}_{1}+(\Delta v_{\bot})(\bm{r}\times\hat{\bm{u}}_{1})
\end{equation}
and its square magnitude is
\begin{equation}
|\Delta\bm{L}|^{2}=\left|\frac{\Delta v_{\parallel}}{v_{1}}\bm{L}_{1}\right|^{2}+\frac{2\Delta v_{\parallel}\Delta v_{\bot}}{v_{1}}\left[\bm{L}_{1}\cdot(\bm{r}\times\hat{\bm{u}}_{1})\right]+\left|(\Delta v_{\bot})(\bm{r}\times\hat{\bm{u}}_{1})\right|^{2}
\end{equation}
when averaging the above expression we note that the middle additive term does not contribute, because $\langle\Delta v_{\parallel}\Delta v_{\bot}\rangle=0$. Therefore the average is simply
\begin{equation}
\left\langle |\Delta\bm{L}|^{2}\right\rangle =\left(\frac{\bm{L}_{1}}{v_{1}}\right)^{2}\left\langle (\Delta v_{\parallel})^{2}\right\rangle +\left\langle |\bm{r}\times\hat{\bm{u}}_{1}|^{2}\right\rangle \left\langle (\Delta v_{\bot})^{2}\right\rangle .
\end{equation}
The vector multiplication in the last term is easy to average, as $\bm{r}$ is a constant vector and $\hat{\bm{u}}_{1}$ is a unity vector at an arbitrary direction with respect to $\bm{r}$, the angle $\gamma$ between them is uniformly distributed, leading to
\begin{equation}
\left\langle |\bm{r}\times\hat{\bm{u}}_{1}|^{2}\right\rangle =r^{2}\left\langle |\sin\gamma|^{2}\right\rangle =\frac{1}{2}r^{2}
\end{equation}
and therefore, also noting that $L=rv_{t}$
\begin{equation}
\left\langle |\Delta\bm{L}|^{2}\right\rangle =\frac{r^{2}}{v^{2}}\left[v_{t}^{2}\left\langle (\Delta v_{\parallel})^{2}\right\rangle +\frac{1}{2}v^{2}\left\langle (\Delta v_{\bot})^{2}\right\rangle \right]\label{eq:deltaL-vector}
\end{equation}
Equation (\ref{eq:deltaL-vector}) above is similar in form to Equation (A10) of Paper I; the difference is that the coefficient of $\langle(\Delta v_{\bot})^{2}\rangle$ within the square brackets is proportional to $v^{2}$ instead of $v_{r}^{2}$.

We finalize the calculation of the diffusion coefficients writing $\langle(\Delta v_{\parallel})^{2}\rangle$ and $\langle(\Delta v_{\bot})^{2}\rangle$ as functions of velocity with the Rosenbluth potentials as substituting into Equation (\ref{eq:deltaL-vector})
\begin{equation}
\frac{\left\langle |\Delta\bm{L}|^{2}\right\rangle }{\Delta t}=\frac{8\pi\Gamma r^{2}}{3v}\left\{ \left(\frac{1}{2}v^{2}-v_{r}^{2}\right)F_{4}(v)+\frac{3}{2}v^{2}F_{2}(v)+\left(2v^{2}-v_{r}^{2}\right)E_{1}(v)\right\} \label{eq:deltaL-Rosenbluth}
\end{equation}
where $\Gamma=4\pi G^{2}m^{2}\ln\Lambda$ and $F_{n}$ and $E_{n}$ are the Rosenbluth potentials. This implies assumption of the isotropy of the distribution function. The calculation proceeds by orbitally averaging Equation (\ref{eq:deltaL-Rosenbluth}) as shown in Appendix B of Paper I. The result is shown in Figure \ref{fig:diffusion}.

\begin{figure}
\begin{center}
\includegraphics[width=0.5\textwidth]{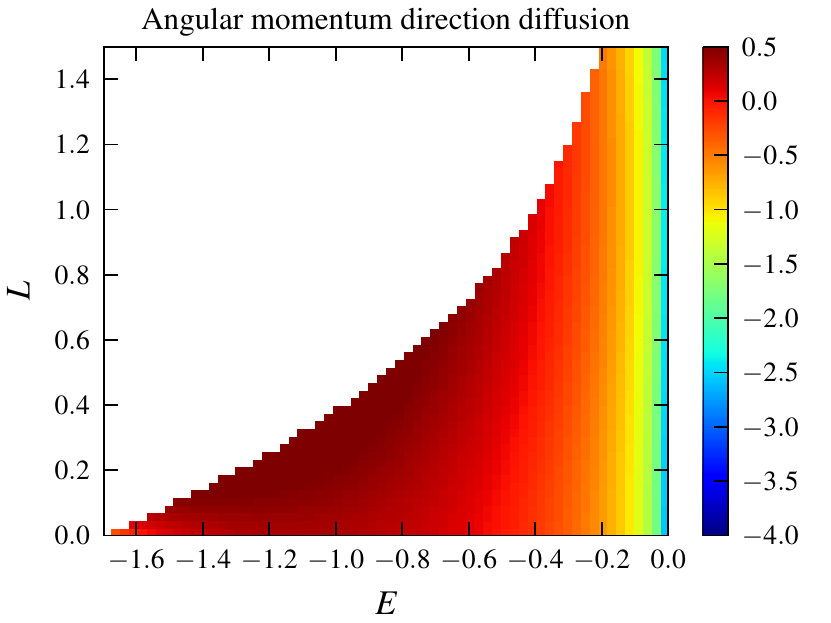}
\end{center}
\caption{Angular momentum direction diffusion coefficient for a Plummer model due to 2-body relaxation only, i.e. neglecting VRR, calculated by orbital averaging scattering theory for each point on a grid in $(E,L)$-space. The color scale is $\log_{10}[ND_{L_\mathrm{v}^{2}}/(\ln\Lambda\langle L^{2}\rangle)]$ where $N$ is the number of particles, $\ln\Lambda$ is the Coulomb logarithm, and $\langle L^{2}\rangle$ is the mean square angular momentum of particles in a Plummer model. This normalization guarantees dimensionlessness and independence of the number of particles. However, note that the RMS angle by which the angular momentum vectors are displaced due to 2-body relaxation is proportional to $(\langle L^2\rangle/L^2)^{1/2} [ND_{L_\mathrm{v}^{2}}/(\ln\Lambda\langle L^{2}\rangle]^{1/2}$. The axes are in Hénon units for a Plummer model (see footnote \ref{fn:units}). \label{fig:diffusion}}
\end{figure}

\section{Angular momentum angular correlation function on the sphere}
\label{app:correlation}
VRR describes the RMS rate at which the angular momentum vectors directions reorient and diffuse on the spherical surface. A useful statistical measure to describe this process beyond $\eta_{\rm v}$ and $\beta_{\rm v}$ defined in equation~(\ref{eq:rms-deltaLv}) is the correlation function. The correlation function describes how a fluctuating field $\rho$ correlates between two points separated by a distance $\Delta r$. In particular in a two-dimensional plane of area $A$ this may be calculated as
\begin{equation}
C(\Delta r) = \int_A \bm{dr}'\int_A \bm{d r}''\,  \rho(\bm{r}')\, \rho(\bm{r}'')\, \delta_{\rm D}(\Delta r - |\bm{r}'-\bm{r}''|)
\end{equation}
where $\delta_{\rm D}(\cdot)$ denotes the $\delta$-function\footnote{the D index stands for Dirac to avoid confusion with other $\delta$ labels in the paper}. On the spherical surface, $S_2$, this may be defined using the proper distance $\alpha = \cos^{-1}({\hat{\bm{r}}'}\cdot {\hat{\bm{r}}''})$ or its cosine, i.e. $\mu = \cos \alpha$
\begin{equation}\label{eq:Cmudef}
C(\mu) = \frac{1}{2\pi}\int_{S_2} d\hat{\bm{r}}'\int_{S_2} d\hat{\bm{r}}''\, \rho(\bm{r}')\, \rho(\bm{r}'') \,\delta_{\rm D}(\mu - \hat{\bm{r}}'\cdot\hat{\bm{r}}''|)\,.
\end{equation}
where the integration is over the unit sphere.

In practice this equation may be evaluated using Legendre-polynomials and spherical harmonics\footnote{ We use the definition which satisfies
\begin{equation}
 \int Y_{\ell m}(r)Y_{\ell' m'}^*(r)\, d \Omega =  \delta_{\ell\, \ell'}\delta_{m\, m'}
\end{equation}
if $\ell\geq 0$ and $-\ell \leq m\leq \ell$, and similarly for $(\ell',m')$. } which satisfy
\begin{align}
&\delta(\mu - \cos \gamma)= \sum_{\ell=0}^{\infty}
\frac{2\ell+1}{2}P_{\ell}(\cos\gamma)\,P_{\ell}(\mu)\,,\\\label{eq:Legedre}
&P_{\ell}(\cos\gamma) =
\frac{4\pi}{2\ell+1}\sum_{m=-\ell}^{\ell}
Y_{\ell m}(\hat{\bm{r}})\, Y_{\ell m}^*(\hat{\bm{r}}')\,
\end{align}
for any $\cos\gamma=\hat{\bm{r}}\cdot \hat{\bm{r}}'$, $\mu$, and $\ell\geq 0$. Substituting in equation~(\ref{eq:Cmudef}) gives
\begin{equation}
C(\mu) = \int_{S_2} d\hat{\bm{r}}'\int_{S_2} d\hat{\bm{r}}'' \rho(\bm{r}')\, \rho(\bm{r}'') \sum_{\ell=0}^{\infty}\sum_{m=-\ell}^{\ell}
Y_{\ell m}(\hat{\bm{r}})\, Y_{\ell m}^*(\hat{\bm{r}}')\,P_{\ell}(\mu)\,.
\end{equation}
Reversing the order of the sums and the integral, this may be written as superposition of Legendre polynomials
\begin{equation}
C(\mu) = \sum_{\ell=0}^{\infty} C_{\ell}  P_{\ell}(\mu)
\end{equation}
where
\begin{align}\label{eq:Cell}
C_{\ell} &= \sum_{m=-\ell}^{\ell} |c_{\ell m}|^2\,,\\
c_{\ell m} &=  \int_{S_2} d\hat{\bm{r}}\, \rho(\hat{\bm{r}})\, Y_{\ell m}(\hat{\bm{r}}) \,.\label{eq:c_ellm0}
\end{align}
In particular, if the density field $\rho(\hat{\bm{r}})$ is given by $N$ discrete particles on the sphere at $\hat{\bm{r}}_i$ for $i\in \{1,\dots,N\}$, we may evaluate the density weighted integral in equation~(\ref{eq:c_ellm0}) as an ensemble average over the particles in the sample:
\begin{equation}
c_{\ell m } = \frac{1}{N}\sum_{i=1}^{N} Y_{\ell m}(\hat{\bm{r}}_i)\,.
\end{equation}
Substituting in equation~\eqref{eq:Cell} and using equation~(\ref{eq:Legedre}) and using $|c_{\ell m}|^2=c_{\ell m}c_{\ell m}^*$ (where the asterisk $^*$ denotes complex conjugate) finally gives
\begin{align}
C_{\ell} &=  \sum_{m=-\ell}^{\ell}\left[\frac{1}{N}\sum_{i=1}^{N} Y_{\ell m}(\bm{r}_i)\right]\left[\frac{1}{N}\sum_{j=1}^{N} Y^*_{\ell m}(\bm{r}_j)\right]  = \frac{1}{N^2} \sum_{i,j=1}^{N} \sum_{m=-\ell}^{\ell}  Y_{\ell m}(\hat{\bm{r}}_i)\, Y_{\ell m}(\hat{\bm{r}}_j) = \frac{2\ell + 1}{4\pi N^2}\sum_{i,j=1}^{N} P_{\ell}(\hat{\bm{r}}_i\cdot \hat{\bm{r}}_j)\,.
\end{align}
This result gives the angular correlation function at any given instant. The correlation function in both angle and time may be derived similarly from equation~(\ref{eq:Cmudef}) but by adding up the contributions of the correlation between $\rho(\bm{r}',t_0)$ and $\rho(\bm{r}'',t_0+\Delta t)$ separated by an angular distance $\mu = \cos \alpha = \bm{r}'\cdot\bm{r}''$, averaged over all such $\bm{r}'$, $\bm{r}''$, and $t_0$. For the distibution of angular momentum vector directions, we get
\begin{equation}
C(\mu,\Delta t) = \sum_{\ell=0}^{\infty} C_{\ell}(\Delta t)\,  P_{\ell}(\mu)\,,\\
\end{equation}
where
\begin{equation}
C_{\ell}(\Delta t) = \sum_{m=-\ell}^{\ell}\left[\frac{1}{N}\sum_{i=1}^{N} Y_{\ell m}[\hat{\bm{L}}_i(t_0)]\right]\left[\frac{1}{N}\sum_{j=1}^{N} Y^*_{\ell m}[\hat{\bm{L}}_j(t_0+\Delta t)]\right] = \frac{2\ell + 1}{4\pi N^2} \sum_{i,j=1}^{N} P_{\ell}\left[\hat{\bm{L}}_i(t_0) \cdot \hat{\bm{L}}_j(t_0+\Delta t) \right]\,,\label{eq:Cell2}
\end{equation}
where averaging is implicitly assumed over the reference time $t_0$. The result shows that the angular correlation function is expressed in the basis of $P_{\ell}(\mu)$ with $C_{\ell}$ coefficients which are independent of the angular scale $\mu$ and which depend on time. The $C_{\ell}$ coefficients specify the power on characteristic angular scales of $\theta \sim \pi/(2\ell)$ when comparing the distribution functions at times separated by $\Delta t$.

An interesting special case to examine is when the angular momenta of different stars in a given zone are uncorrelated with each other, e.g. if each star samples an isotropic distribution independently at every $t_0$ instant. In this case one can show that only the $i=j$ terms contribute to the sum over $i$ and $j$ in equation~(\ref{eq:Cell2})\footnote{we adjust the normalization to remain $N$-independent}:
\begin{align}
C_{\ell}(\Delta t) &= \frac{1}{N}\sum_{m=-\ell}^{\ell} \sum_{i=1}^{N}
Y_{\ell m}[\hat{\bm{L}}_i(t_0)]\, Y^*_{\ell m}[\hat{\bm{L}}_i(t_0+\Delta t)] = \frac{2\ell + 1}{4\pi N} \sum_{i=1}^{N}
P_{\ell}\left[\hat{\bm{L}}_i(t_0) \cdot \hat{\bm{L}}_i(t_0+\Delta t) \right]\,.\label{eq:Cell}
\end{align}
Note that at $\Delta t=0$, the dot products are unity and $P_{\ell}(1)=1$ for all $\ell$, and so $C(\mu,0)=\delta_{\rm D}(\mu)$. For larger $\Delta t$ as the dot products decrease, all Legendre polynomials decrease, and hence $C(\mu,\Delta t)$ decreases for $\mu=1$. Once the particles forget their initial conditions $C_{\ell}\approx \rm const$ up to a shot noise type fluctuation. Note, that if we did not neglect the cross-correlation among different particles and used equation~(\ref{eq:Cell2}), then $C(\mu,0)$ would be nonzero for $\mu>0$.

We may express the angular correlation function equivalently with the angular variance $V_{\ell}$ defined next. For Brownian motion on the sphere with diffusion coefficient $D$, the spherical moments follow \citep{Kocsis+15}
\begin{equation}
\langle Y_{\ell m}\rangle = e^{-\frac14\ell(\ell+1) V_{\ell} }
\end{equation}
where $V_{\ell} = D \Delta t$ is the angular variance that represents the mean squared angle that the particle has moved during Brownian motion in time $t$. Substituting in equation~(\ref{eq:Cell}) shows how the angular correlation function changes for Brownian motion:
\begin{equation}\label{eq:Brownian}
C_{\ell} (\Delta t) = \frac{2\ell + 1}{4\pi} e^{-\frac14\ell(\ell+1) V_{\ell} }\,.
\end{equation}
While $V_{\ell}= D \Delta t$ for all $\ell$ for Brownian motion, these coefficients may be different for an arbitrary random process. Given an arbitrary angular correlation function $C_{\ell}(\Delta t)$, the angular variance is defined by solving equation~(\ref{eq:Brownian}) for $V_{\ell}$:
\begin{equation}
V_{\ell}(\Delta t) = \frac{-4}{\ell(\ell+1)}\ln \left[ \frac{4\pi}{2\ell +1} C_{\ell}(\Delta t)  \right] = \frac{-4}{\ell(\ell+1)}\ln \frac{1}{N}\sum_{i=1}^{N}P_{\ell}\left[\hat{\bm{L}}_i(t_0) \cdot \hat{\bm{L}}_i(t_0+\Delta t) \right]\,,
\end{equation}
In the last line we substituted equation~(\ref{eq:Cell}) for $C_{\ell}(\Delta t)$. Note that averaging is assumed implicitly over $t_0$. For a nuclear star cluster bound to a massive object $M$, \citet{Kocsis+15} have shown that the angular momentum vector directions change as
\begin{align}
V_{\ell}&\approx \eta_{{\rm 2body},\Omega} \frac{Nm}{M}  \tau \quad \text{random walk}\,,\\
V_{\ell}&\approx \beta_{{\rm coherent\ VRR},\Omega}^2 \frac{Nm}{M}\tau^2 \quad \text{constant torque}\,,\\
V_{\ell}&\approx \beta_{{\rm incoherent\ VRR},\Omega} \frac{Nm}{M}  \tau \quad \text{random walk}\,,\\
V_{\ell}&\approx \text{saturated fluctuations around } \frac{2\ln[(2\ell+1)N T]}{\ell(\ell +1)}
\end{align}
where $\beta_{{\rm coherent\ VRR},\Omega}=\beta_{\rm v} L_\mathrm{c}/L$ and similarly for $\eta_{{\rm 2body},\Omega}$, $N$ denotes the selected number of particles on the angular momentum shell (e.g. region-II) and $T$ denotes the number of analyzed independent time segments with different $t_0$ start times. The $V_{\ell}$ coefficients of angular momentum vector directions grow initially linearly with time $\tau$ due to 2-body relaxation, then gradually change slope to a quadratic function due to coherent VRR, then change back to a linear function of $\tau$ during incoherent VRR until the particles forget their initial conditions and the completely uncorrelated (fully mixed) state is reached.

\end{document}